\documentclass[reprint,superscriptaddress,amsmath,amssymb,aps,pre,hyphens,floatfix]{revtex4-1}
\usepackage{graphicx}
\usepackage{dcolumn}
\usepackage{bm}
\usepackage{hyperref}
\usepackage{xcolor}
\usepackage{comment}
\usepackage{csquotes}
\usepackage{epstopdf}
\usepackage{subfigure}
\usepackage{natbib}

\begin{document}

% submitted to Phys. Rev. E (2020); arXiv:2008.11541

\title{Critical Dynamics of Anisotropic Antiferromagnets in an External Field}
\author{Riya Nandi} \email{riya11@vt.edu}
\affiliation{Department of Physics and Center for Soft Matter and Biological Physics, \\
	Virginia Tech, Blacksburg, VA 24061, USA}
\author{Uwe C. T\"auber} \email{tauber@vt.edu}
\affiliation{Department of Physics and Center for Soft Matter and Biological Physics, \\ 
	Virginia Tech, Blacksburg, VA 24061, USA}

\date{\today} % {Received: date / Accepted: date}

\begin{abstract}
We numerically investigate the non-equilibrium critical dynamics in three-dimensional 
anisotropic antiferromagnets in the presence of an external magnetic field. 
The phase diagram of this system exhibits two critical lines that meet at a bicritical point. 
The non-conserved components of the staggered magnetization order parameter couple 
dynamically to the conserved component of the magnetization density along the 
direction of the external field. 
Employing a hybrid computational algorithm that combines reversible spin precession 
with relaxational Monte Carlo updates, we study the aging scaling dynamics for the 
model C critical line, identifying the critical initial slip, autocorrelation, and aging 
exponents for both the order parameter and conserved field, thus also verifying the 
dynamic critical exponent. 
We further probe the model F critical line by investigating the system size dependence of
the characteristic spin wave frequencies near criticality, and measure the dynamic critical
exponents for the order parameter including its aging scaling at the bicritical point. 
\end{abstract}


\maketitle

\section{Introduction}
\label{intro}

Classical anisotropic antiferromagnets governed by Heisenberg spin exchange in an
external magnetic field exhibit a rich phase diagram with multiple thermodynamic
ground states separated by continuous and discontinuous transition lines that meet at a 
multicritical point~\cite{kosterlitz1976bicritical, liu1973quantum}. 
This paradigmatic model system describes various magnetic compounds including
$\mathrm{MnF_2}$~\cite{shapira1970magnetic}, 
$\mathrm{GdAlO_3}$~\cite{rohrer1977bicritical}, 
$\mathrm{NiCl_2\cdot 6H_2O}$~\cite{oliveira1978magnetic}, and thus has been
extensively investigated theoretically as well as experimentally. 
Early renormalization group~\cite{fisher1974spin}, Monte Carlo 
simulation~\cite{landau1978phase}, and high-temperature 
expansion~\cite{mouritsen1980general} studies have systematically explored its 
complex phase diagram and characterized the properties of the different ordered phases 
and determined the universality classes for its critical transition lines. 

\begin{figure}
    \centering
    \includegraphics[width=8cm]{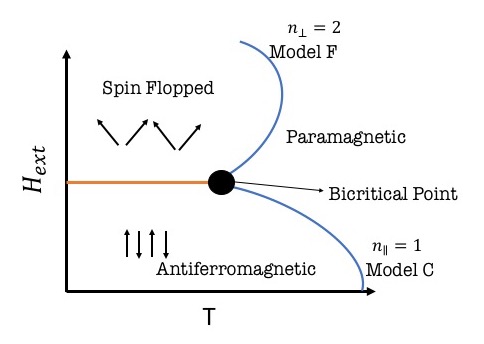}
    \caption{Schematic $H_{\rm ext}$-$T$ phase diagram of an anisotropic
    antiferromagnet in the presence of an external field showing two ordered phases 
    separated by a first-order transition line. 
    The antiferromagnetic and spin-flopped ordered phases are separated from the 
    disordered paramagnetic phase by lines of continuous phase transitions that meet 
    at a bicritical point.}
    \label{pd}
\end{figure}

The anisotropy term along one of the crystal axes breaks the rotational symmetry of the
Heisenberg antiferromagnet enforcing an antiparallel spin ordering along that axis in the 
low-temperature, low-field ground state, i.e., the Ising antiferromagnetic phase (AF), 
c.f.~Fig.~\ref{pd}. 
As the external field strength is tuned up while keeping the temperature low, the ground
state switches, via a discontinuous (first-order) transition, to a spin-flop phase (SF). 
At higher values of either temperature or magnetic field, the system becomes 
paramagnetic (PM). 
The phase transitions between the PM and the AF and SF states are both continuous 
(second-order).
While the associated static critical properties of the system are characterized solely by
the symmetry and the dimensionality of the anisotropic Heisenberg Hamiltonian, the
dynamics in the vicinity of the phase transitions driving it from the disordered phase to
the ordered AF and SF phases, respectively, are distinctly different and crucially depend
on the microscopic, reversible dynamical couplings between the corresponding order 
parameters and the conserved magnetization components. 

Indeed, the transition from the antiferromagnetic to the paramagnetic phase is 
described by the dynamic critical behavior of model C, while the spin-flop to 
paramagnetic phase transition belongs to the dynamical universality class of model F; 
here we invoke the classification introduced by Halperin and Hohenberg in 1977 in their 
comprehensive early review of dynamic critical 
phenomena~\cite{hohenberg1977theory}. 
The presence of a non-ordering field shifts the non-universal parameters of the system 
such as the critical temperature, but it does not change the nature of the critical point,
i.e, the universal scaling exponents characterizing the critical power law divergences
remain unaltered. 
Yet an intriguing distinct physical scenario results in the vicinity of the
\textit{multicritical point}, where both critical lines as well as the discontinuous phase
transition meet. 
At the multicritical point, all three different phases compete for the lowest free energy 
configuration; thus the system's symmetry at this special point is higher than in the 
adjacent parameter space. 
Hence, the dynamical properties in the vicinity of such a point are expected to be 
characterized by new exponents which are distinct from those of the conjoining critical 
lines. 

The nature of the multicritical point for anisotropic antiferromagnets subject to an
external magnetic field (oriented along the $z$ direction) has been somewhat 
controversial. 
This special point in parameter space is characterized by two coupled order parameter 
fields with $O(n_\perp) \bigoplus O(n_\parallel)$ symmetry; it displays long-range
order both along the magnetic field $z$ axis and in the perpendicular $xy$ plane. 
A series of papers by Folk, Holovatch, and Moser employed the renormalization group 
to investigate the static and dynamic critical behavior and the stability of the associated 
fixed points of the system~\cite{folk2008field, folk2008field2, folk2009field, folk2012field}.
They predicted the emergence of either bicritical behavior associated with a Heisenberg 
fixed point or tetracritical behavior, in turn associated with a biconical or a decoupled
renormalization group fixed point. 
An analysis to higher order resulted in the tetracritical point with a biconical phase to
represent the stable fixed point~\cite{calabrese2003multicritical}. 
A recent six-loop perturbative dimensional $\epsilon$ expansion of the three-dimensional 
$n$-vector model has shown that the for $n=3$, the obtained critical exponents are very 
close to the Heisenberg fixed point values even though that is not the stable fixed point in 
the asymptotic infrared limit~\cite{adzhemyan2019six}. 
Thus, the observed critical behavior in experiments and simulations may be 
indistinguishable from the isotropic Heisenberg fixed point scaling. 
This has indeed been observed to be true in a series of detailed Monte Carlo simulations 
analyzing order parameter susceptibilities, the Binder cumulant, and associated probability 
distributions, which reported that the nature of the multicritical point is in fact 
\textit{bicritical} with Heisenberg symmetry~\cite{selke2011evidence, tsai2014bicritical}. 
% Subsequently, a full renormalization group study of the flow equations at the 
% multicritical point taking into account the the reversible terms concluded that in 
% two-loop order the biconical fixed point becomes stable, whereas the Heisenberg fixed 
% remains stable in a one-loop calculation~\cite{folk2012field}. 
% In contrast, a series of detailed Monte Carlo simulations analyzing order parameter 
% susceptibilities, the Binder cumulant, and associated probability distributions provided 
% concrete evidence that the nature of the multicritical point is in fact \textit{bicritical} with 
% Heisenberg symmetry~\cite{selke2011evidence, tsai2014bicritical}.

While the static critical behavior and stationary critical dynamics of this system has been
investigated comprehensively in the literature, we are not aware of previous 
computational work addressing the non-equilibrium critical relaxation of the anisotropic
Heisenberg antiferromagnet in an external magnetic field near either continuous phase 
transition line, nor at the multicritical point. 
To this end, the system is initially prepared in at a disordered configuration and then 
quenched precisely to its critical point such that the dynamics algebraically slowly 
evolves towards the asymptotic stationary state. 
During this early non-equilibrium relaxation time window, the system retains the 
memory of its initial state, and thus manifests broken time translation invariance. 
By studying the ensuing aging scaling behavior of two-time quantities at these early 
times, one may fully characterize the dynamics near the distinct critical points and the 
corresponding universality classes~\cite{henkel2010non, tauber2014critical}.

In this work, we utilize a hybrid simulation method combining relaxational Monte Carlo
kinetics and reversible spin precession processes~\cite{nandi2019nonuniversal} to
explore the aging scaling behavior of the AF-to-PM model C critical line, and in the
vicinity of the multicritical point. 
We measure the critical aging scaling, autocorrelation decay, and initial slip exponents 
for the model C universality class, and determine the associated dynamic critical 
exponent. 
We also investigate the non-equilibrium dynamics of the conserved magnetization
component. 
Furthermore, we utilize Fourier spectral analysis for the spin wave excitations in the 
$xy$ plane in the spin-flop ordered phase to verify the dynamic critical exponent for the
SF-to-PM model F critical line.
Finally, we investigate the critical order parameter dynamics at the bicritical point.

\section{Model and Simulation Method} 

The Hamiltonian of a three-dimensional antiferromagnet with anisotropic Heisenberg 
exchange interactions in an external magnetic field is given by
\begin{equation}
    \mathcal{H}= J \sum_{\langle ij \rangle}^N \left[ \Delta(S^x_iS^x_j+S^y_iS^y_j) 
    + S^z_iS^z_j \right] - H_{\rm ext} \sum_{i=0}^N S^z_i \, ,
    \label{hamiltonian}
\end{equation}
where $S_x^i$, $S_y^i$, $S_z^i$ represent the components of the three-dimensional 
spin vector $\vec{S}^i$ at the $i$th site of a simple cubic lattice of linear extension $L$ 
with total spin number $N = L^3$. 
The magnitude of the spins are fixed to unit magnitude, 
$S_x^{i2}+S_y^{i2}+S_z^{i2}=1$. 
$J > 0$ denotes the antiferromagnetic exchange interaction along the $z$ axis between 
nearest-neighbor spin pairs $\langle ij \rangle$; we set $J = +1$, i.e., measure 
temperature in units of $J / k_{\rm B}$ and the external field in units of $J$. 
The uniaxial anisotropy $0 < \Delta < 1$ imposes an ``easy'' $z$ axis such that the 
spins would order anti-parallel along this direction in the absence of an external 
field~\cite{selke2011evidence}; we choose $\Delta = 0.8$.
The presence of this anisotropy explicitly breaks the $O(3)$ rotational symmetry of the 
Hamiltonian in spin space and splits it into two subspaces of dimensions 
$n_{\parallel} = 1$ and $n_{\perp} = 2$. 
Thus its static critical properties are governed by the universality class with 
$O(1) \bigoplus O(2)$ symmetry with associated Fisher exponent 
$\eta \approx 0.04$~\cite{folk2008field}. 

Applying an external magnetic field $H_{\rm ext} \ne 0$ along the $z$ axis forces the 
uniaxially aligned spins to flop over into the $xy$ plane beyond some critical field 
strength $H_{\rm ext}^c$. 
Thus, at low temperature $T$ and external field $H_{\rm ext}$, i.e. in the AF phase, 
the $z$ component of the staggered magnetization 
\begin{equation}
    \phi_{\parallel} = \sum_{i,j,k=0}^L (-1)^{p} S_{i,j,k}^z
    \label{staggz}
\end{equation}
represents the non-conserved order parameter for the system; here the indices $(i,j,k)$ 
denote the three spatial directions, and $p = i+j+k$ ensures that the sum extends over 
the differences between every alternate spin in the lattice. 
Upon increasing $H_{\rm ext}$, the value of $\phi_{\parallel}$ is diminished, and 
instead the staggered magnetization components in the $xy$ plane perpendicular to the 
applied field become appreciable.
Thus in the SF phase, an appropriate, also non-conserved, order parameter is a 
two-component vector $\vec{\phi}_{\perp} = (\phi_x, \phi_y)$ with magnitude
\begin{equation}
   \phi_{\perp} = \sqrt{\biggl( \sum_{i,j,k=0}^L (-1)^p S_{i,j,k}^x \biggr)^2 
   + \biggl( \sum_{i,j,k=0}^L (-1)^p S_{i,j,k}^y \biggr)^2} \ .
   \label{staggxy}
\end{equation}

It is important to note that the Hamiltonian (\ref{hamiltonian}) conserves the $z$ 
component of the total magnetization 
\begin{equation}
    M_z = \displaystyle \sum_{i,j,k=0}^N S_{i,j,k}^z 
    \label{magnz}
\end{equation}
under the dynamics, $\{ \mathcal{H} , M_z \} = 0$; here the Poisson bracket 
constitutes the classical counterpart of the quantum-mechanical commutation relation 
between the spin angular momentum and the Hamiltonian. 
The dynamical mode couplings between conserved magnetization fluctuations and the 
order parameter components decisively influence the antiferromagnet's critical 
dynamics~\cite{freedman1976critical, hohenberg1977theory, folk2006review, 
tauber2014critical}. 
Indeed, in addition to the irreversible, relaxational terms arising from the static 
couplings in the Hamiltonian, one must account for the \textit{reversible} kinetics 
caused by the underlying microscopic dynamics between the order parameter and any
conserved modes. 
At zero temperature, the microscopic equations of motion obeyed by each spin variable 
are $d S_i^\alpha(t)/dt = \{ \mathcal{H} , S_i^\alpha(t) \}$, where the spin vector 
components satisfy the standard angular momentum Poisson brackets 
$\{ S_i^\alpha , S_j^\beta \} = \sum_\gamma \epsilon^{\alpha\beta\gamma} \, 
S_i^\gamma \, \delta_{ij}$ with the fully antisymmetric unit tensor 
$\epsilon^{\alpha\beta\gamma} = \pm 1$. 

In the $n_{\parallel} = 1$ subspace $\{M_z,\phi_{\parallel}\}=0$; thus there is no 
reversible coupling term. 
However, in the $n_{\perp} = 2$ subspace, the non-conserved vector order parameter 
$\vec{\phi}_{\perp}$ couples reversibly to the conserved magnetization (\ref{magnz})
\begin{equation}
     \{M_z,\phi_{\alpha}\}=\epsilon_{\alpha\beta z}\phi_{\beta} \, ,
     \label{precession}
\end{equation}
where $\alpha, \beta \in \{ x,y \}$. 
This non-vanishing mode coupling gives rise to the following deterministic equations of 
motion of the microscopic spin components at 
$T = 0$~\cite{ma1974critical, freedman1976critical}
\begin{equation}
     \frac{d\vec{S}_i(t)}{dt} = 
     \vec{S}_i(t) \times \frac{\partial \mathcal{H}}{\partial \vec{S}_i(t)} \ ,
     \label{rungekutta}
 \end{equation}
which describe precession of the unit spin vector in the local effective field.

To simulate the dynamics of this system at finite temperature, one needs to implement 
relaxation terms as well as the reversible microscopic equations of motion. 
For convenience, we work with the two angular degrees of freedom $\vartheta$ and 
$\varphi$ that are related to the unit vector spin components through $(S_x,S_y,S_z) =
(\sin \vartheta \cos \varphi, \sin \vartheta \sin \varphi, \cos \vartheta)$.
To respect the underlying conservation property, the azimuthal angles $\vartheta$ are 
updated by means of Kawasaki Monte Carlo kinetics where two randomly picked 
neighboring spins exchange their $\vartheta$ values following the standard Metropolis 
rules~\cite{kawasaki1966diffusion}. 
In contrast, the polar angles $\varphi$ are evolved using Glauber dynamics where the 
spin component at the selected lattice site is subjected to a finite rotation with again 
Metropolis updates~\cite{glauber1963time}. 
These Monte Carlo update steps of the spin configurations are alternated with a 
fourth-order Runge-Kutta integration of the equations of motion (\ref{rungekutta})
\cite{nandi2019nonuniversal}. 
For our simulation, the integration was performed in parallel on all spins over discrete 
time increments $\Delta t = 0.01 / J$ with each integration step separated by $10$ 
Monte Carlo sweeps over the entire lattice. 
We determined this combination to be optimal in maintaining the conservation laws 
within the truncation error bounds of the numerical integration scheme.

\section{Model C dynamical scaling}

The dynamical universality class conventionally labeled as model C describes the pure 
relaxation dynamics of a non-conserved $n$-component critical order parameter field, 
coupled to a conserved density~\cite{hohenberg1977theory, 
akkineni2004nonequilibrium, folk2006review, tauber2014critical}.
In the present study, the low-temperature, low-field ground state is an Ising
antiferromagnet; hence $n=1$ with the $z$-component of the magnetization density 
constituting the conserved scalar field $m$. 
A two-loop renormalization group calculation demonstrated that the scalar model C 
($n=1$) is governed by a strong-scaling fixed point with both the order parameter 
relaxation and the conserved density diffusion scaling with the same anomalous 
exponent $\alpha / \nu$~\cite{folk2004critical}. 
This results in the \textit{dynamic critical exponent} 
$z_z = z_m = 2 + \alpha / \nu \approx 2.185$; here 
$\nu \approx 0.72$~\cite{campostrini2002critical,landau1994computer} 
describes the algebraic divergence of the correlation length $\xi \sim |\tau|^{- \nu}$
($\tau \sim T - T_c$), and $\alpha$ is the specific heat critical exponent, 
$C \sim |\tau|^{- \alpha}$, which can be obtained from the hyperscaling relation 
$\alpha = 2 - d \nu$ in $d$ dimensions. 

\begin{figure*}[ht]
    \centering
   \subfigure[]{ \includegraphics[width=\columnwidth]{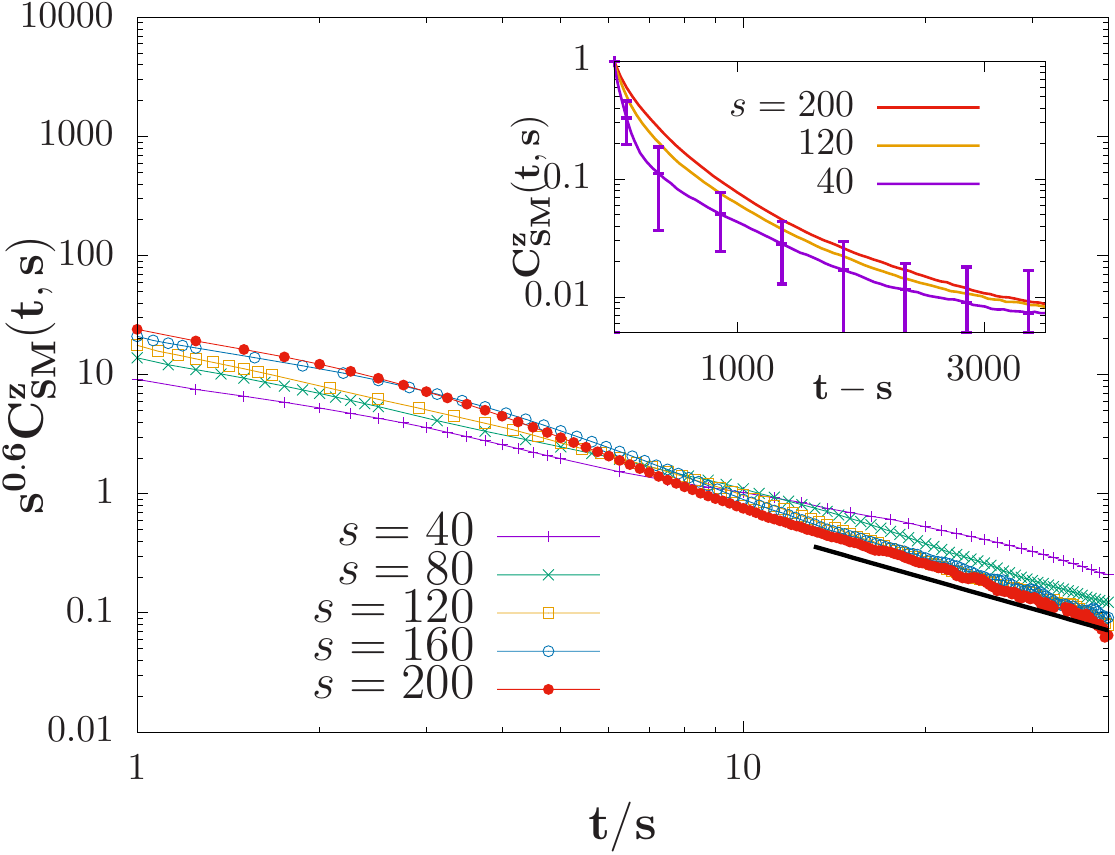} }\hfil
   \subfigure[]{ \includegraphics[width=0.97\columnwidth]{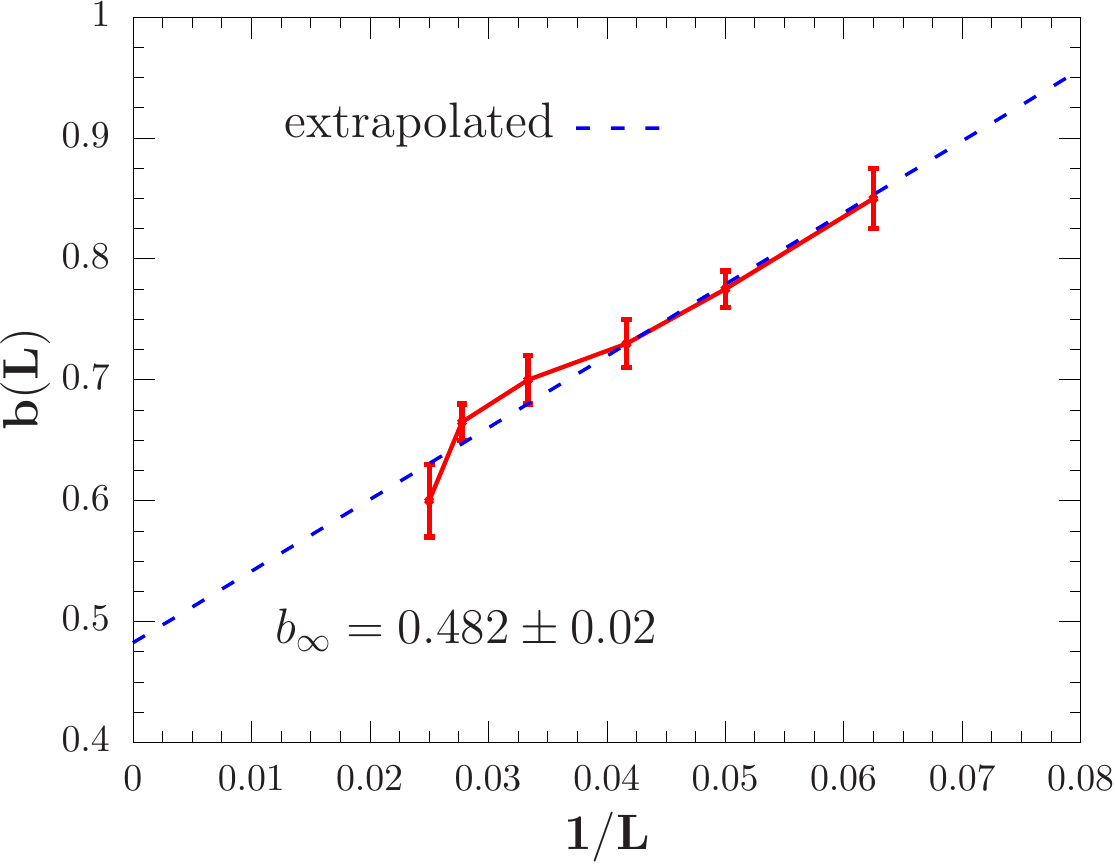}}
    \caption{(a) Aging scaling plots for the two-time spin autocorrelation function
    $C_{\rm SM}^z(t,s)$ of the Ising antiferromagnetic order parameter on a simple 
    cubic lattice of linear system size $L = 40$ with periodic boundary conditions. 
    The system is quenched from an initially disordered configuration to the critical point
    at $T_c = 1.5$, $H_{\rm ext}^c = 3.0$. 
    Double-logarithmic rescaled graphs for different waiting times $s$ collapse with the
    scaling exponent $b=0.6$. 
    The inset shows the autocorrelation plots as a function of $t - s$ (in simulation time 
    steps) for $s = 200, 120, 40$ (top to bottom), demonstrating broken time translation
    invariance. 
    Statistical errors are indicated in the graph for the shortest waiting time $s = 40$. 
    (b) Finite-size extrapolation analysis for the aging exponent $b(L)$ plotted vs. $1/L$ for 
    six different system sizes $L = 16 \ldots 40$. 
    A linear extrapolation to the infinite system size limit $L \to \infty$ yields 
    $b_\infty=0.482 \pm 0.02$.}
    \label{modelc}
\end{figure*}
The non-equilibrium relaxation of model C was investigated by Oerding and Janssen 
using the dynamic renormalization group approach~\cite{oerding1993nonequilibrium}.
Following a critical quench, the two-time order parameter correlation function relating
two space-time points at distance $r$ and times $s < t$ satisfies the scaling law 
\begin{equation}
    C(t,s,r,\tau) = r^{- (d - 2 + \eta)} \, (t / s)^{\theta - 1} \ {\hat C}(r / \xi, t / \xi^z)
    \, ,
    \label{agingsc}
\end{equation}
where $\theta$ is the \textit{initial slip exponent} representing a new independent 
universal exponent for purely dissipative systems with non-conserved order 
parameter~\cite{janssen1989new}. 
It also describes the power law growth of the order parameter in the early-time  
universal regime which sets in right after the microscopic time during the 
non-equilibrium relaxation process.
At the critical temperature $T = T_c$ ($\tau = 0$), the two-time autocorrelation 
function ($r = 0$) assumes the simple-aging scaling form~\cite{henkel2010non}, 
\begin{equation}
    C(t,s) \sim s^{-b} \, (t / s)^{-\lambda / z} \, ,
    \label{aging}
\end{equation}
with the static and dynamic exponents related to the \textit{scaling collapse exponent}
$b$ via
\begin{equation}
    b = (d - 2 + \eta) / z \, ,
    \label{b}
\end{equation} 
and to the \textit{autocorrelation exponent} $\lambda$ according to
\begin{equation}
    \lambda = d - 2 + \eta + z (1 - \theta) = z (1 + b - \theta) \, .
    \label{lambda}
\end{equation} 
For model C with $n=1$, a second-order perturbative renormalization calculation 
predicts $\theta \approx 0.27$ in three dimensions, if one boldly extrapolates the 
dimensional expansion in $\epsilon = 4 - \epsilon$ to 
$\epsilon = 1$~\cite{oerding1993nonequilibrium}.  

\begin{figure}[hb]
    \centering
    \includegraphics[width=\columnwidth]{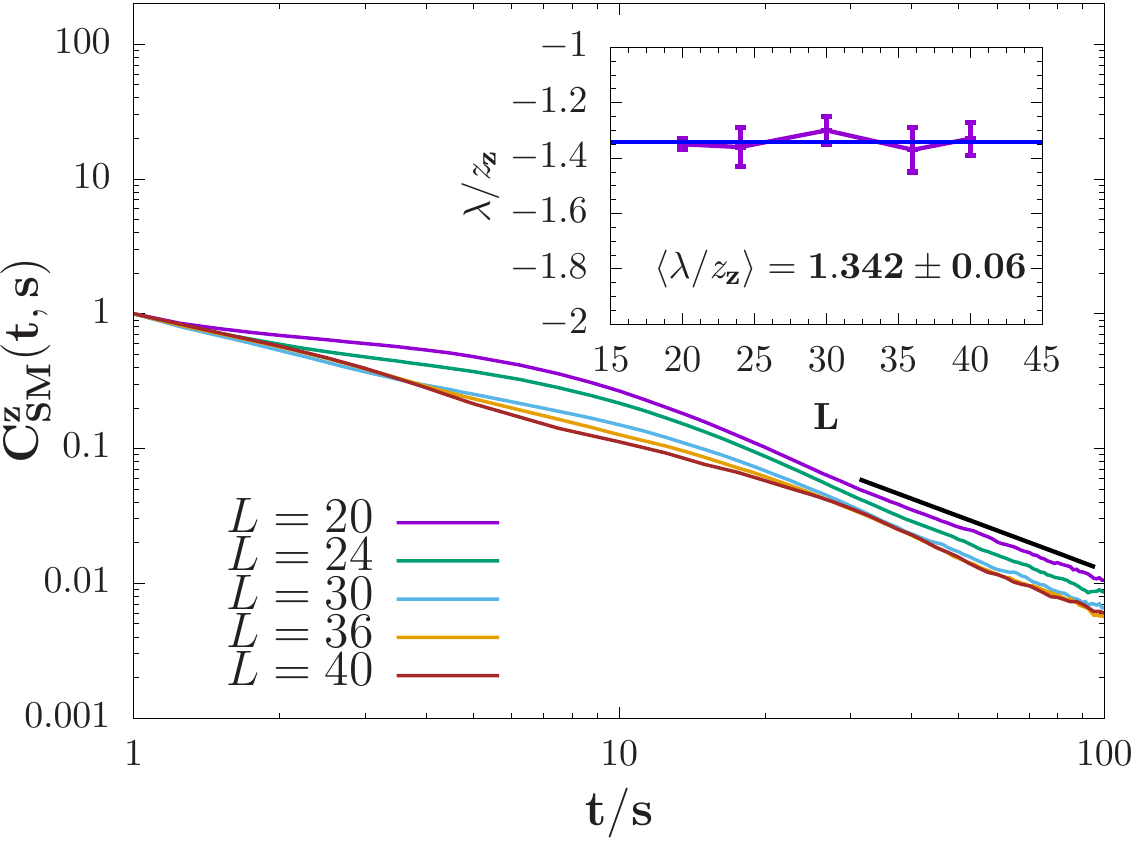}
    \caption{Double-logarithmic plots of the two-time staggered magnetization 
    autocorrelation function vs. $t / s$ for different linear system sizes $L=20,\ldots, 40$ 
    (top to bottom) taken at an early waiting time $s = 40$ exhibit power law decays in
    the long-time limit $t \gg s$. 
    The mean value of the autocorrelation exponent is determined to be 
    $\langle \lambda / z_z \rangle = 1.342 \pm 0.06$.}
    \label{thetac}
\end{figure} 
\begin{figure*}[ht]
    \centering
   \subfigure[]{ \includegraphics[width=\columnwidth]{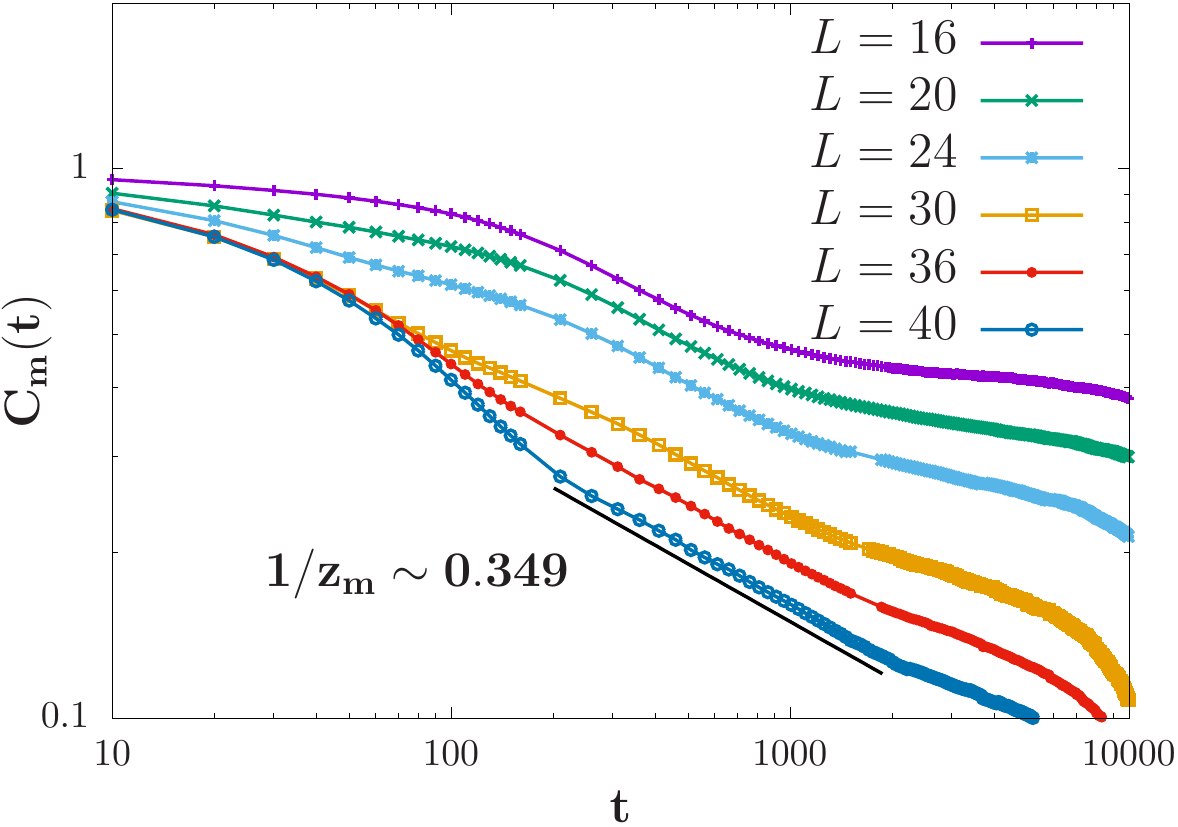}} \hfill
   \subfigure[]{ \includegraphics[width=0.98\columnwidth]{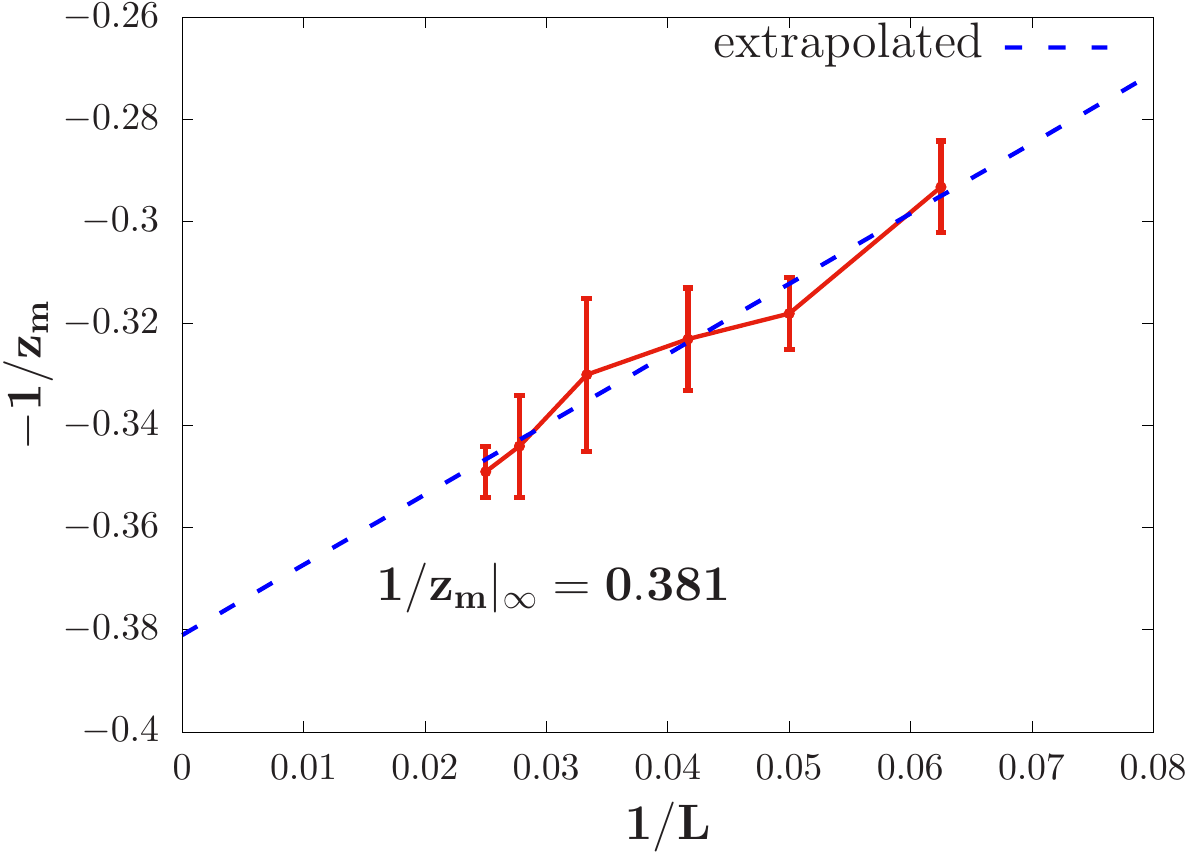}}
    \caption{(a) Double-logarithmic plots of the autocorrelation function for the 
    conserved magnetization component $M_z$ vs. time $t$ for different linear system 
    sizes $L$ display an intermediate regime governed by algebraic decay. 
    The system was quenched to the critical point at $T_c = 1.5$, 
    $H_{\rm ext}^c =3.0$. 
    (b) Finite-size extrapolation analysis for the decay exponent plotted vs. $1 / L$ for six 
    different system sizes $L=16 \ldots 40$. 
    A linear extrapolation to the infinite system size limit $L \to \infty$  yields
    $1/z_m = 0.381$.}
    \label{modelcmz}
\end{figure*}
To numerically study the critical relaxation and the aging scaling regime we initialized 
the system in a disordered spin orientation configuration corresponding to a very high 
temperature, and subsequently performed critical quenches to a point ($T_c = 1.5$, 
$H_{\rm ext}^c = 3.0$) on the model C critical line. 
As shown in Fig.~\ref{modelc}(a), we obtain an aging scaling window for waiting times 
$s = 120 \ldots 200$ simulation time steps. 
The inset demonstrates that time translation invariance is broken, as the two-time 
correlation function for the order parameter $C_{\rm SM}^z(t,s)$ is not simply a
function of the time difference $t - s$ as it would be in the stationary limit, but evolves 
differently for the distinct waiting times $s$ in this temporal window. 
By collapsing the data of $C_{\rm SM}^z(t,s)$ for several waiting times $s$ plotted as 
a function of the time ratio $t / s$ in accordance with Eq.~(\ref{aging}), one can obtain
the collapse exponent $b$ for which we find $b \approx 0.6$ for linear system 
extension $L = 40$. 
The data collapse is noticeably improved for both later waiting times $s$ and larger 
observation times $t$. 
This is expected since the simple-aging scaling form (\ref{aging}) is supposed to hold 
only for sufficiently large $t \gg s$ and $s$. 

However, in our finite simulation domain, data for large times are inevitably hindered by 
finite-size effects. 
Thus, in order to better estimate the asymptotic collapse exponent, we perform a 
systematic finite-size extrapolation analysis by plotting $b$ vs. $1 / L$ for system sizes 
$L=16, 20, 24, 30, 36, 40$, c.f.~Fig.~\ref{modelc}(b). 
Linear extrapolation to infinite system size $L \rightarrow \infty$ leads to 
$b_\infty = 0.482 \pm 0.02$. 
We then obtain from Eq.~(\ref{b}) the dynamic exponent for the order parameter in 
model C in $d = 3$ dimensions, $z_z = (1 + \eta) / b =  2.158 \pm 0.09$. 
This result agrees well within our errors with the theoretically predicted value
$z_z \approx 2.185$. 
The autocorrelation exponent $\lambda / z_z$ can be extracted from the power law tails 
of $C_{\rm SM}^z(t,s)$, apparent in double-logarithmic plots vs. $t / s$ for times 
$t \gg s$.
Fig.~\ref{thetac} displays the data for five different system sizes, from which we obtain 
the mean value $\langle \lambda / z_z \rangle = 1.342 \pm 0.06$. 
Using Eq.~(\ref{lambda}), one may obtain the initial slip exponent 
$\theta = 0.14 \pm 0.08$. 
This value shows a similar trend as the theoretical prediction, but unsurprisingly, differs 
in magnitude by about a factor $2$ from the naive extrapolation of the second-order
results of the perturbative $\epsilon$ expansion about the mean-field value $\theta = 0$.
\begin{figure}[hb]
    \centering
    \includegraphics[width=\columnwidth]{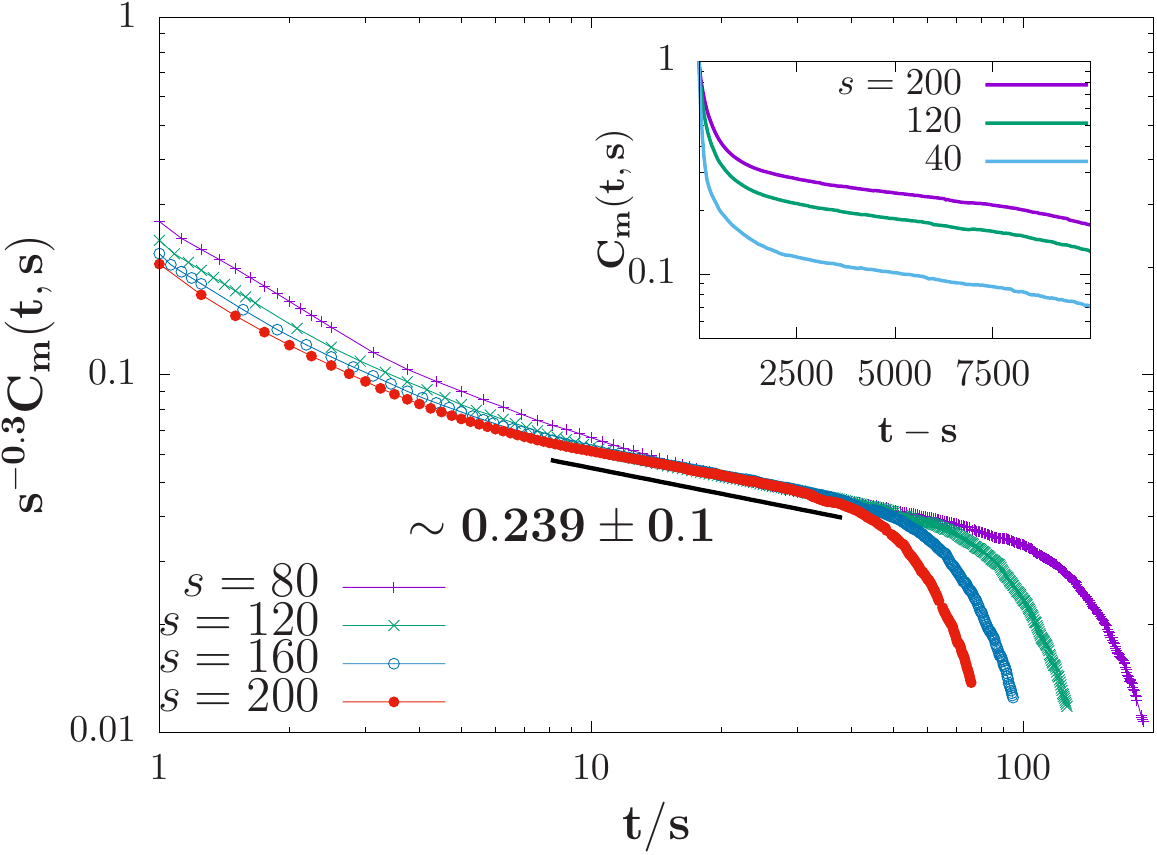}
    \caption{Aging scaling plots for the two-time autocorrelation function $C_m(t,s)$ of 
    the conserved magnetization for linear system size $L = 40$. 
    Double-logarithmic rescaled plots for different waiting times $s$
    collapses with exponent $b_m \approx - 0.3$ and decay exponent 
    $ \approx 0.239 \pm 0.1$.
    The inset shows the autocorrelation plots as a function of $t - s$ for $s = 200, 120, 40$
    (top to bottom), demonstrating broken time translation invariance.}
    \label{agingmz}
\end{figure}

\begin{figure*}[ht]
    \centering
   \subfigure[]{ \includegraphics[width=\columnwidth]{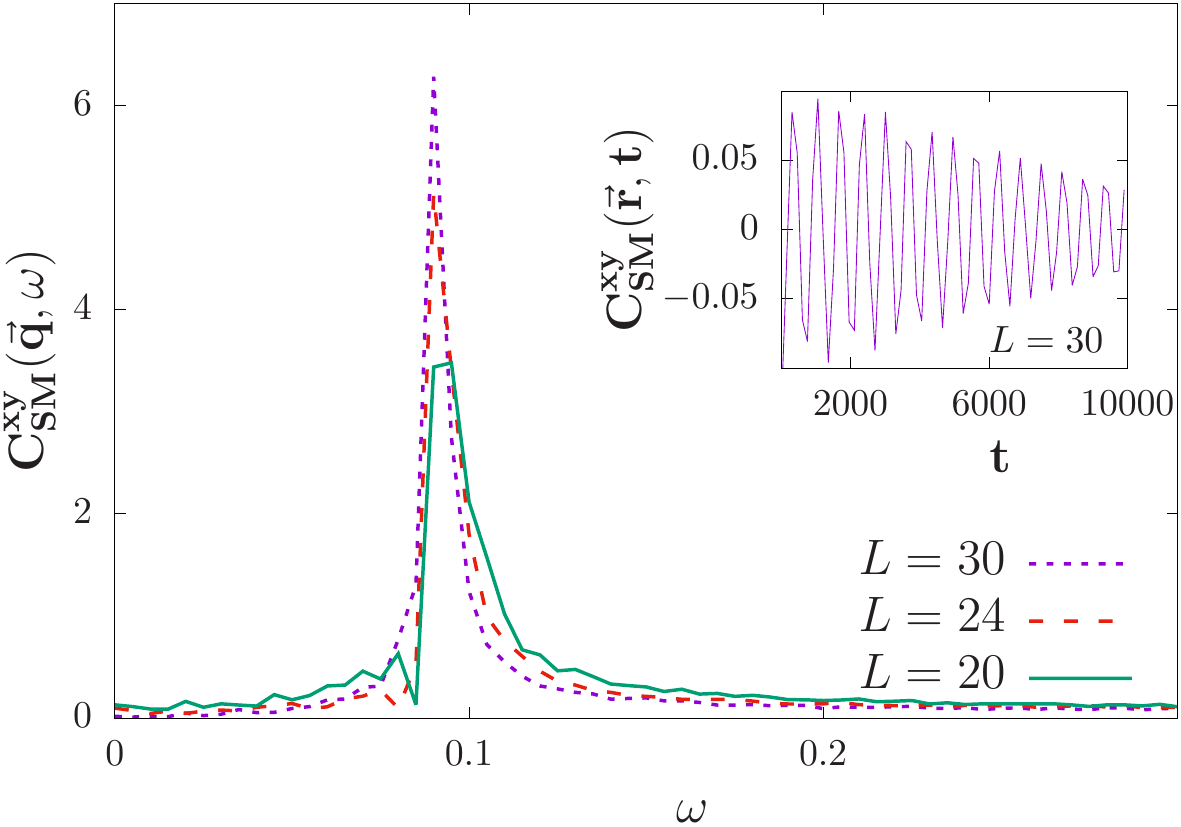}} \hfill
  \subfigure[]{  \includegraphics[width=\columnwidth]{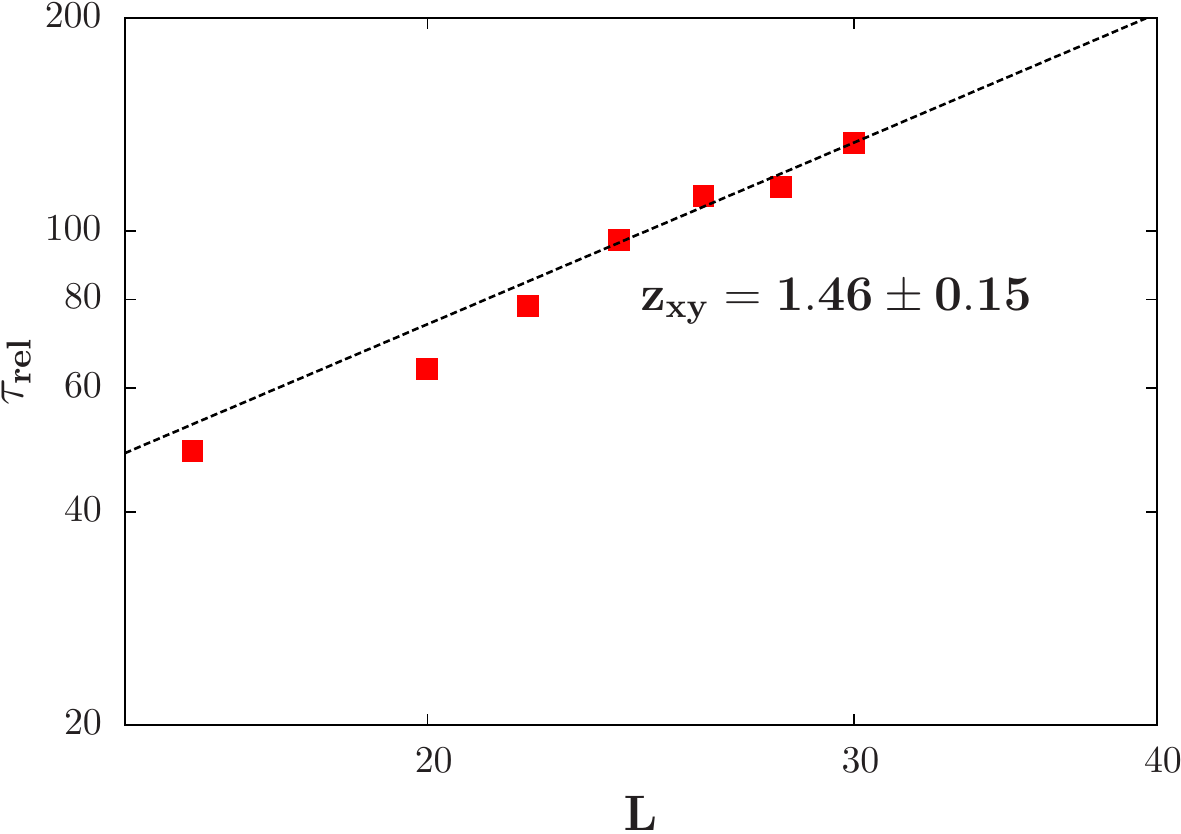}}
    \caption{(a) Fourier spectrum of the staggered magnetization correlation function in 
    the $xy$ plane for different linear system sizes $L = 20, 24, 30$ for a non-zero wave 
    vector $\vec{q}$ vs. frequency $\omega$. 
    The inset shows propagating spin waves in the spin-spin correlation function with 
    spatial separation $\vec{r}$ as a function of time. 
    (b) Double-logarithmic plot of the relaxation time $\tau_{\rm rel}$ vs. linear system 
    sizes $L=16 \ldots 32$. 
    Fitting the data for large $L > 20$ yields the dynamic critical exponent 
    $z = 1.46 \pm 0.15$.}
    \label{fig:modelf}
\end{figure*}
We further explore the non-equilibrium dynamics of the conserved magnetization 
component $M_z$ which is reversibly coupled to the ordered parameter. 
This non-critical conserved field undergoes diffusive relaxation with correlations
$C_m(\vec{q},\omega)\sim q^{-2}\hat{C}(\omega/q^z)$ or equivalently 
$C_m(\vec{r},t) \sim r^{-(d-2)}\tilde{C}(r/t^z)$. 
The asymptotic long-time scaling form for the temporal magnetization correlation function
at the critical temperature thus becomes
\begin{equation}
    C_m(t) \sim t^{- (d - 2) / z} \, . 
    \label{mzC}
\end{equation} 
In the data depicted in Fig.~\ref{modelcmz}(a), we discern an intermediate power law 
region in the decay of the spin autocorrelation function before the graphs fall off 
exponentially due to finite-size effects. 
As expected, this algebraic regime becomes more prominent upon increasing the linear 
system size $L$. 
We again perform a systematic finite-size extrapolation to obtain the asymptotic value of 
the decay exponent and find $(1 / z_m)_\infty \approx 0.381$, and thus infer 
$z_m \approx 2.62 \pm 0.01$. 
This value however differs from the order parameter dynamic exponent 
$z_z \approx 2.148 \pm 0.1$, indicating that likely the time-scale ratio between the 
staggered magnetization relaxation and magnetization diffusion still has not reached its 
asymptotic fixed point, and the strong dynamic scaling hypothesis cannot be validated. 
In this context, we direct the readers towards previous work by Koch and Dohm discussing 
the effect of finite system size on the relaxation and diffusion time scales of 
model C~\cite{koch1998finite}.

One can also obtain the aging scaling data from the two-time autocorrelation function for the 
conserved magnetization, c.f.~Fig.~\ref{agingmz}. 
We note though that for large waiting times we observe another power-law region at later 
times which is distinctly different from the previously obtained algebraic decay in the 
intermediate relaxation regime of the single-time autocorrelation function. 
It is at these later times that the rescaled plots for different waiting times collapse with an 
exponent $b_m \sim -0.3$ and a decay exponent equal to $0.239 \pm 0.01$. 
Earlier analyses of conserved spin systems have predicted two regimes with different power 
laws, with a new length scale governing the crossover between both algebraic 
regimes~\cite{sire2004autocorrelation, godreche2004non}. 
Ultimately in the long-time limit, however, the decay of the autocorrelations is determined by 
only one length scale, independent of the waiting time $s$. 
In a similar vein, the non-critical conserved magnetization here displays the signature of two 
distinct scaling regimes. 
Yet unlike in the conserved spin systems, we observe an early relaxation regime with a faster 
power law decay, prominent in the single-time autocorrelation plots in Fig.~\ref{modelcmz}, 
which subsequently crosses over to a slower algebraic decay until ultimately finite-size effects 
dominate. 
Moreover, the negative value of the exponent $b_m$ suggests the presence of long-lived 
metastable states. 
The precise nature of the crossover scaling for the conserved magnetization in our system
thus remains open for future investigation.

\section{Model F dynamical scaling}

The continuous phase transition between the spin flop and the paramagnetic phases is 
described by the dynamic universality class model 
F~\cite{hohenberg1977theory, folk2006review}. 
Also known as the \enquote{asymmetric planar spin 
model}~\cite{halperin1976renormalization}, this universality class describes the critical
dynamics of a two-component vector order parameter coupled reversibly to a conserved 
scalar density in the presence of an external $Z_2$ symmetry-breaking field. 
The only other known and prominent physical system described by model F is the 
normal- to superfluid transition in $^4 \textrm{He}$~\cite{josephson1966relation}.  
In anisotropic antiferromagnets, the non-conserved components of the planar staggered 
magnetization $\phi_x$ and $\phi_y$ couple reversibly through the non-vanishing 
Poisson brackets to the conserved magnetization component $M_z$, resulting in the 
precession motion (\ref{precession}) of the spin vectors around a local field produced 
by their exchange interaction with their nearest neighbors and the external field.
\begin{figure*}[ht]
   \centering
   \subfigure[]{ \includegraphics[width=\columnwidth]{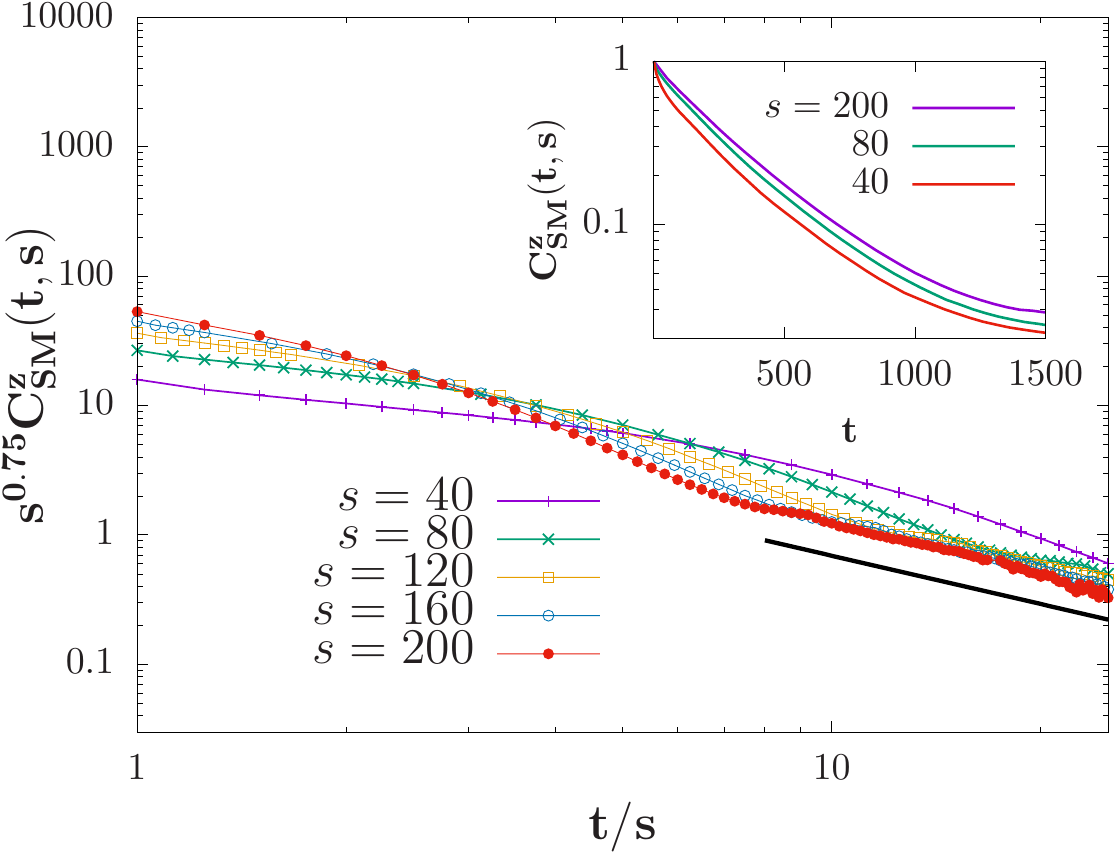} }\hfil
   \subfigure[]{ \includegraphics[width=0.97\columnwidth]{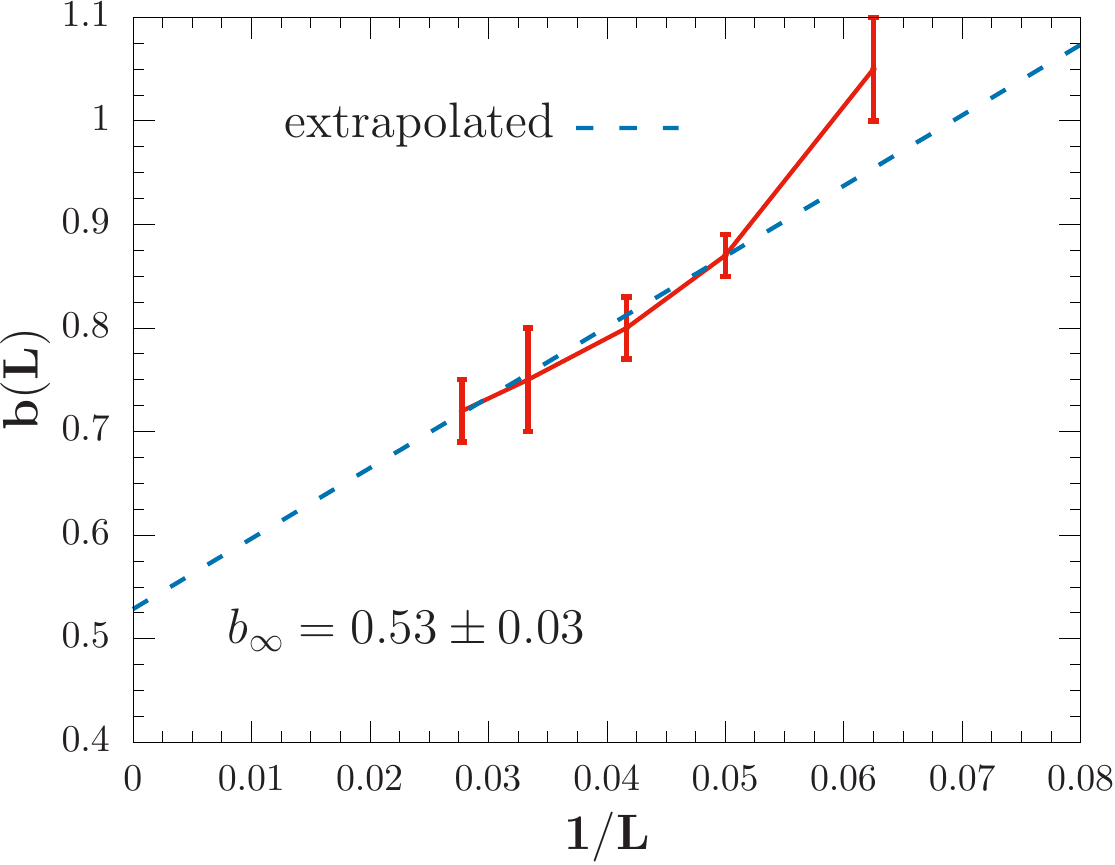}}
    \caption{(a) Aging scaling plots for two-time spin autocorrelation function
    $C_{\rm SM}^z(t,s)$ of the Ising antiferromagnetic order parameter on a simple 
    cubic lattice of linear system size $L=30$ with periodic boundary conditions. 
    The system is quenched from an initially disordered configuration to the bicritical 
    point at $T_c = 1.025$, $H_{\rm ext}^c = 3.825$. 
    Double-logarithmic rescaled graphs for different waiting times $s$ collapse with the
    scaling exponent $b \approx 0.75$. 
    The inset shows the autocorrelation plots as a function of $t - s$ for 
    $s = 200, 80, 40$ (top to bottom), demonstrating broken time translation invariance. 
    (b) Finite-size extrapolation analysis for the aging exponent $b$ plotted vs. $1/L$ 
    for five different linear system sizes $L = 16 \ldots 36$. 
    A linear extrapolation using the data from the four largest systems to the infinite 
    system size limit $L \to \infty$ yields $b_\infty=0.53 \pm 0.03$.}
   \label{bicb}
\end{figure*}

One may view the conserved magnetization components acting as the infinitesimal 
rotation generators for the order parameter components, resulting in propagating spin 
waves in the ordered phase~\cite{halperin1969hydrodynamic, tauber2014critical}. 
The spin wave damping $\Gamma_c$ decreases as the critical temperature is 
approached, with the associated relaxation time $\tau_{\rm rel} = 1 / \Gamma_c$ 
diverging at $T_c$. 
Hence procuring the aging scaling data by probing the conventional two-time 
correlations turns out not to be a viable approach at the model F critical line.
Moreover, owing to the reversible mode couplings between the conserved magnetization 
and the non-conserved order parameter components, the initial slip exponent $\theta$ 
and hence the autocorrelation exponent $\lambda$ are expected to be non-universal in 
this case; specifically, these exponents should depend on the initial distribution of the 
magnitudes of the conserved modes~\cite{nandi2019nonuniversal,oerding1993non}. 

However, one can extract the dynamic exponent from the temporal evolution of the 
stationary correlation function in the vicinity of the critical parameters in the ordered 
phase.
Near the critical temperature, the spin wave oscillations have an exponentially 
decreasing amplitude $\sim e^{- \Gamma_c t} \sim e^{- t / \tau_{\rm rel}}$. 
In a finite system near $T_c$, the stationary relaxation time diverges with linear system 
size with the dynamic critical exponent $z$ characterizing its \textit{critical slowing 
down}: $\tau_{\rm rel} \sim L^z$. 
We have obtained the relaxation time via measuring the half-peak width $\Gamma_c$ 
of the Fourier transform of the spin-spin correlation function~\cite{chen2016non}, 
$C_{\rm SM}^{xy}(\vec{q},\omega)=\int C_{\rm SM}^{xy}(\vec{r},t) e^{i\omega t} 
dt$, see Fig.~\ref{fig:modelf}(a). 
The asymptotic value of the dynamic exponent for model F is known exactly from the 
dynamic renormalization group, $z_{xy} = d / 2$ in $d \leq 4$
dimensions~\cite{halperin1976renormalization,gunton1976renormalization}.
From our relaxation time data as function linear system size, we find that for the five 
largest $L$ the best fit line gives $z_{xy} \approx 1.46$, within our error bars
in agreement with the theoretical predition $1.5$, c.f.~Fig.~\ref{fig:modelf}(b). 
As one would expect, with larger system sizes $z_{xy}$ tends towards the asymptotic 
value.

\section{Bicritical dynamical scaling}

\begin{figure}[hb]
    \centering
    \includegraphics[width=\columnwidth]{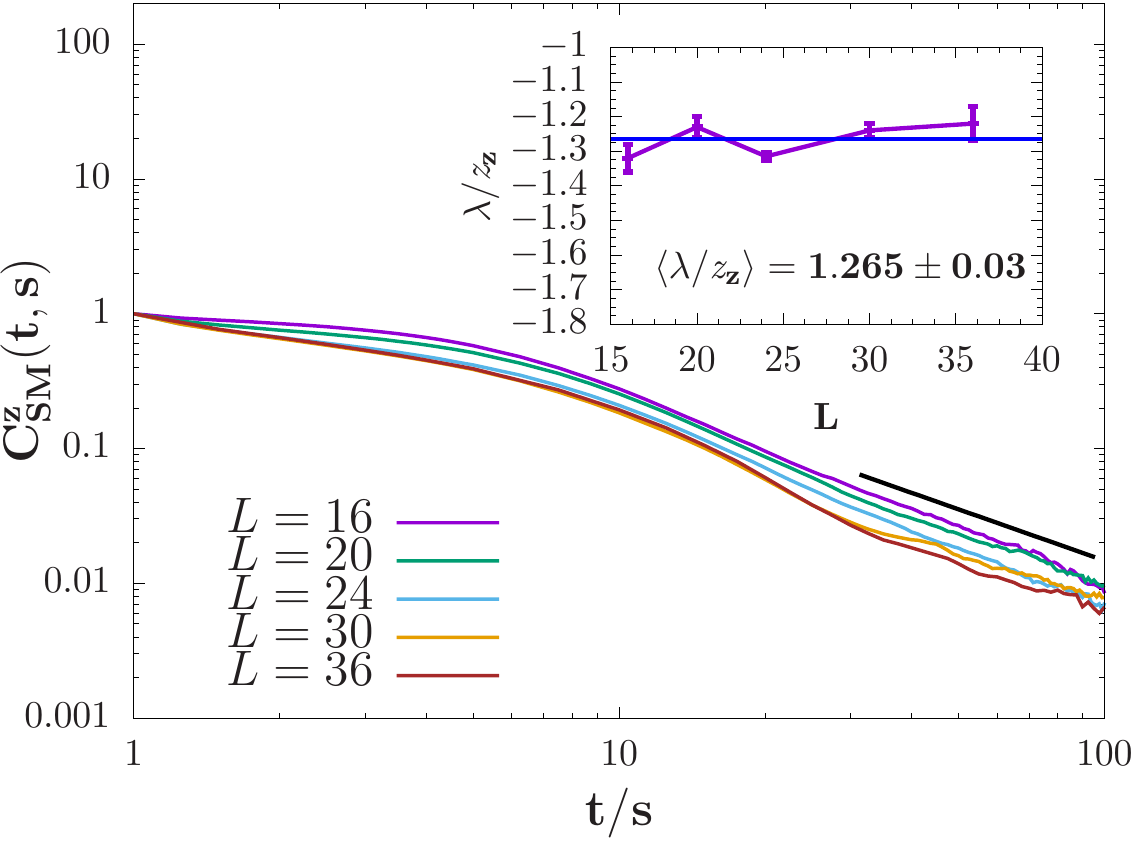}
    \caption{Double-logarithmic plots of the two-time staggered magnetization 
    autocorrelation function vs. $t / s$ for different linear system sizes $L=16,\ldots, 36$ 
    (top to bottom) taken at an early waiting time $s = 40$ exhibit power law decays in 
    the long-time limit $t \gg s$. 
    The mean value of the autocorrelation exponent is determined to be 
    $\langle \lambda / z_z \rangle = 1.265 \pm 0.03$.}
    \label{bic}
\end{figure}
The two continuous phase transition lines described by models C and F meet at a 
\textit{bicritical point} which is described by a different dynamical universality class.
In their field-theoretical analysis, Folk, Holovatch, and Moser found that irrespective of 
whether the static behavior of the system is described by the Heisenberg or biconical 
renormalization group fixed point, the parallel and perpendicular order parameter
components scale similarly in time with dynamic critical exponents 
$z_z \sim z_{xy} \approx 2.003$ and $z_m \approx 1.542$ in the asymptotic 
limit~\cite{folk2012field}. 
However, strong non-asymptotic effects originating from the mode coupling terms in 
the vicinity of the bicritical point lead to very different crossover dynamical exponents 
which exhibit weak dynamical scaling. 

Similar to the model C analysis, we obtain a dynamic aging scaling window for waiting 
times $s = 80 \ldots 200$ for the easy-axis scalar order parameter at the bicritical point. 
Figure~\ref{bicb} depicts the scaling collapse of the two-time autocorrelation plots for 
different waiting times for a system with linear size $L = 30$ with aging exponent 
$b \approx 0.75$.
A subsequent system size extrapolation yields the asymptotic value 
$b_\infty = 0.53 \pm 0.03$. 
Using Eq.~(\ref{b}), one can then infer the dynamic critical exponent 
$z_z = 1.962 \pm 0.15$ which is in agreement with the theoretical prediction within 
our error bars. 
From the mean value of $\langle \lambda / z_z \rangle = 1.265 \pm 0.03$ over five 
different system sizes (c.f.~Fig.~\ref{bic}), we also obtain the bicritical initial slip 
exponent $\theta =0.265 \pm 0.05$ for the staggered magnetization along the $z$ axis.

\section{Conclusion}

We have utilized a hybrid numerical method that incorporates reversible spin precession 
dynamics through a deterministic integration scheme with relaxational Monte Carlo kinetics 
to investigate the both the stationary critical dynamics and the non-equilibrium critical 
relaxation in three-dimensional anisotropic antiferromagnets in an external magnetic field. 
From the aging scaling data of the order parameter spin autocorrelation function at the 
model C critical line, we obtained the aging and autocorrelation exponents. 
A systematic finite-size extrapolation analysis allowed for the extraction of the asymptotic 
value of the aging collapse exponent $b \approx 0.482$, which leads to the dynamic 
exponent $z_z \approx 2.148$. 
This is in very good agreement with the theoretical prediction. 
Further, we report the value of the initial slip exponent $\theta \approx 0.14$. 
Additionally, we extract the dynamic exponent for the conserved magnetization 
$z_m \approx 2.62$ and observe two distinct time scales in the decay of its two-time spin 
autocorrelation function.

In the vicinity of the model F critical line, the presence of spin waves hindered the aging 
scaling analysis. 
However, from the Fourier transform analysis of the spin waves we obtained the critical 
relaxation times that increase algebraically with system size, with the dynamic critical 
exponent $z_{xy} \approx 1.46$, which agrees with the theoretical value $z_{xy} = 3/2$ 
within our systematic and statistical errors.  
Finally, we performed an aging scaling analysis for the scalar order parameter component 
along the direction of the external field at the bicritical point. 
We thus verified that the dynamic exponent at this point $z_z \approx 1.962$ is different 
from the corresponding values at both the model C and model F critical lines, contrasting 
the nature of the dynamical universality class at the bicritical point. 

\section*{Acknowledgments}
We would like to thank Michel Pleimling for fruitful discussions and valuable suggestions 
that aided us in this project. 
Research was sponsored by the U.S. Army Research Office and was accomplished under 
Grant Number W911NF-17-1-0156. 
The views and conclusions contained in this document are those of the authors and should 
not be interpreted as representing the official policies, either expressed or implied, of the 
Army Research Office or the U.S. Government. 
The U.S. Government is authorized to reproduce and distribute reprints for Government 
purposes notwithstanding any copyright notation herein.

\bibliographystyle{spphys} 
\bibliography{bibliography}
\end{document}